\documentclass[aps,twocolumn,showpacs,superscriptaddress,amssymb]{revtex4}
\date{\today}

\usepackage{graphicx,amssymb}
\usepackage{threeparttable}

\begin{document}

\title{Effects of high order deformation on superheavy high-$K$ isomers}

\author{H.L. Liu}
\affiliation{Department of Physics and Astronomy, Texas A\&M
University-Commerce, Commerce, Texas 75429-3011, USA}
\author{F.R. Xu}
\affiliation{School of Physics, Peking University, Beijing 100871,
China}
\author{P.M. Walker}
\affiliation{Department of Physics, University of Surrey,
Guildford, Surrey GU2 7XH, UK}
\author{C.A. Bertulani}
\affiliation{Department of Physics and Astronomy, Texas A\&M
University-Commerce, Commerce, Texas 75429-3011, USA}

\begin{abstract}
Using, for the first time, configuration-constrained
potential-energy-surface calculations with the inclusion of
$\beta_6$ deformation, we find remarkable effects of the high
order deformation on the high-$K$ isomers in $^{254}$No, the focus
of recent spectroscopy experiments on superheavy nuclei. For
shapes with multipolarity six, the isomers are more tightly bound
and, microscopically, have enhanced deformed shell gaps at $N=152$
and $Z=100$. The inclusion of $\beta_6$ deformation significantly
improves the description of the very heavy high-$K$ isomers.
\end{abstract}

\pacs{21.10.-k, 21.60.-n, 23.20.Lv, 27.90.+b}

\maketitle

By overcoming the strong Coulomb repulsion between the large
number of protons, shell effects can lead to the so-called
``island of stability'' centered on a doubly magic nucleus beyond
$^{208}$Pb that has yet to be identified. On the way to the
predicted island, new chemical elements up to
$Z=118$~\cite{HofRMP00,OgaJPG07} have been synthesized, while the
transfermium nuclei have been studied in detail through
spectroscopy experiments~\cite{HerzPPNP08}. Of special note in
spectroscopy studies are multi-quasiparticle (multi-qp) high-$K$
($K$ is the total angular momentum projection onto the symmetry
axis) isomers whose decay to low-$K$ states is inhibited due to
$K$ forbiddenness~\cite{WalNature99}. They provide a probe into
the underlying single-particle structure around the Fermi surface.
For example, the systematic observation of $K^\pi=8^-$ isomers in
$A\approx250$ nuclei demonstrates the existence of $N=152$ and
$Z=100$ deformed shell gaps~\cite{GrePRC08}. Such information is
vital for determining the nuclear potential that can then be used
to predict properties of superheavy nuclei. Furthermore,
superheavy high-$K$ isomers can have enhanced stability against
$\alpha$ decay and spontaneous fission due to unpaired
nucleons~\cite{XuPRL04}, perhaps serving as stepping stones
towards the ``island of stability''.

Among the $A\approx250$ nuclei in which high-$K$ isomers have been
discovered, $^{254}$No has been the focus of recent experiments
due to its relatively high production rate. Two-qp and four-qp
high-$K$ isomers were first established by Herzberg {\it et
al.}~\cite{HerzNature06}, Tandel {\it et al.}~\cite{TanPRL06} and
Kondev {\it et al.}~\cite{KonND07}. Later these isomers were
extensively studied by He\ss{}berger {\it et
al.}~\cite{HessEPJA10} and Clark {\it et al.}~\cite{ClarkPLB10},
with emphasis on the spectrum above the two-qp isomer. All the
experiments agree on the existence of a four-qp isomer with a
half-life in the region of 200 $\mu$s, but the suggested
configurations are controversial. He\ss{}berger {\it et
al.}~\cite{HessEPJA10} and Clark {\it et al.}~\cite{ClarkPLB10}
derived different levels bridging the four-qp and two-qp isomers.
More work is required, both experimental and theoretical, to
confirm the $^{254}$No high-spin level structure.

Theoretical descriptions of superheavy nuclei have made continuous
progress~\cite{SobPPNP07} along with experiments. One important
finding is that high order deformation, especially $\beta_6$, is
significant in modeling very heavy
nuclei~\cite{PatNPA91,PatPLB91}. The inclusion of $\beta_6$
deformation can give extra binding energy in excess of 1 MeV,
resulting in improved reproduction of experimental
masses~\cite{PatNPA91}. The $^{254}$No moment of inertia
calculated with the addition of $\beta_6$ deformation is 17\%
larger than the calculation with only $\beta_2$ and $\beta_4$
deformations~\cite{MunPLB01}. Remarkable $\beta_6$ deformations
were predicted in the $A\approx250$ mass region, with the largest
magnitude ($\beta_6\approx-0.05$) in $^{254}$No~\cite{MunPLB01}.
In this work, we investigate the high order deformation effects on
$^{254}$No high-$K$ isomers.

Configuration-constrained potential-energy-surface (PES)
calculations~\cite{XuPLB98} have been applied to the
three-dimensional deformation space ($\beta_2$, $\beta_4$,
$\beta_6$) to determine the deformations and excitation energies
of multi-qp states. Other frequently-used deformation degrees of
freedom such as $\gamma$ and $\beta_3$ are excluded as they are
calculated to be negligible in $^{254}$No. The observation of
large hindrance in $K$-forbidden $\gamma$-ray transitions (that
indicates approximately good $K$ quantum numbers) in $^{254}$No
has confirmed that the nucleus is well deformed and axially
symmetric~\cite{ClarkPLB10}. Reflection asymmetry can
significantly reduce the outer barrier beyond the second potential
well of a prolate superheavy nucleus, but does not affect the
first well~\cite{MunPLB04}. In addition, $^{254}$No has no
indication of $\beta_8$ deformation~\cite{MunPLB01}. Deformations
with multipolarity higher than eight have been demonstrated to be
negligible in calculations~\cite{PatPLB91}. Therefore, it is
justified for us to limit the calculations to the ($\beta_2$,
$\beta_4$, $\beta_6$) deformation space.

We employ the axially deformed Woods-Saxon potential with the set
of universal parameters~\cite{NazNPA85} to provide single-particle
levels. In order to reduce the unphysical fluctuation of the
weakened pairing field (due to the blocking effect of unpaired
nucleons) an approximate particle-number projection has been used
by means of the Lipkin-Nogami method~\cite{LipNPA73}, with pairing
strengths determined by the average gap method~\cite{MolNPA92}. In
the configuration-constrained PES calculation, it is required to
adiabatically block the unpaired nucleon orbits that specify a
given configuration. This has been achieved by calculating and
identifying the average Nilsson quantum numbers for every orbit
involved in a configuration~\cite{XuPLB98}. The good quantum
numbers of parity and $\Omega$ (the individual angular momentum
projection onto the symmetry axis) facilitate the configuration
constraint in ($\beta_2$, $\beta_4$, $\beta_6$) deformation space.
The total energy of a state consists of a macroscopic part that is
obtained with the standard liquid-drop model~\cite{MyeNP66} and a
microscopic part that is calculated by the Strutinsky
shell-correction approach, including blocking effects. The
configuration-constrained PES calculation can properly treat the
shape polarization due to unpaired nucleons.

\begin{figure}
\includegraphics[scale=0.58]{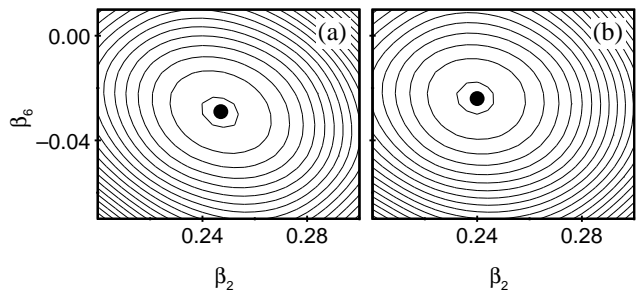}
\caption{\label{fig1}Calculated PESs for $^{254}$No ground state
(a) and
$K^\pi=16^+\{\nu9/2^-[734]\otimes\nu7/2^+[613]\otimes\pi7/2^-
[514]\otimes\pi9/2^+[624]\}$ state (b). At each point
($\beta_2,\beta_6$), the energy is minimized with respect to
$\beta_4$. The energy interval between neighboring contours is 200
keV.}
\end{figure}

\begin{figure}
\includegraphics[scale=0.47]{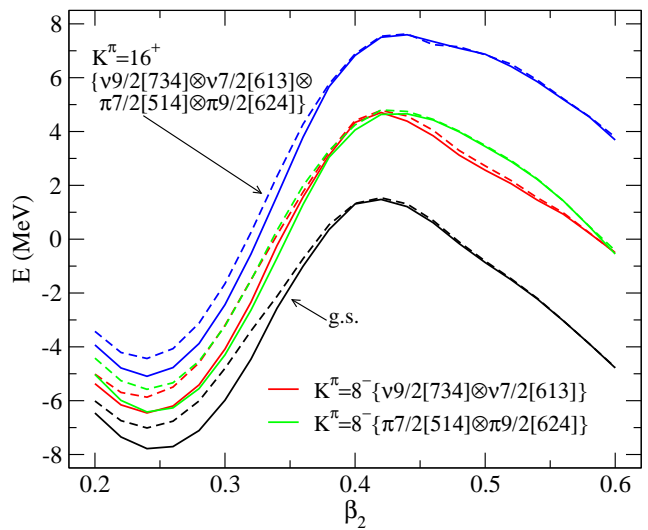}
\caption{\label{fig2}(Color online) $^{254}$No potential energy
curves calculated with (solid lines) and without (dashed lines)
$\beta_6$ deformation. The energy for each $\beta_2$ point is
minimized with respect to deformations $\beta_4$ and $\beta_6$.}
\end{figure}

In Fig.~\ref{fig1}, we display the calculated PESs for $^{254}$No
ground state (g.s.) and four-qp high-$K$ state relevant
to experiments (see below). The PESs show that the states have
remarkable $\beta_6$ deformations. The g.s. $\beta_6$ deformation
-0.029 is smaller in magnitude than -0.05 that was calculated by
Muntian {\it et al.}~\cite{MunPLB01}. This is because we employ
the standard liquid-drop model with a sharp surface for the
macroscopic energy, while Muntian {\it et al.}~\cite{MunPLB01}
used the Yukawa-plus-exponential model with a diffuse surface that
is relatively soft against deformation. Since the latter treatment
seems more realistic, our calculations may slightly underestimate
the magnitude of the $\beta_6$ deformation and hence its effects.
Fig.~\ref{fig1} also shows that the shape of $^{254}$No is robust
against multi-qp excitations, which verifies that the increase in
moment of inertia of the high-$K$ bands with respect to the g.s.
band is due to the reduction of pairing rather than a change of
deformation~\cite{ClarkPLB10}. The influence of the high order
deformation on the stability is significant. The g.s. obtains an
extra binding energy of 0.8~MeV due to $\beta_6$ deformation. The
multi-qp high-$K$ states also have deeper potential wells than
those calculated without $\beta_6$ deformation, as shown in
Fig.~\ref{fig2}. The depth increase for the
$K^\pi=8^-\{\pi7/2^-[514]\otimes\pi9/2^+[624]\}$ state reaches
0.856~MeV. Importantly, our calculations indicate that the
$\beta_6$ deformation has no influence on the barrier peaks (see
Fig.~\ref{fig2}), so that the extra binding energy results in a
net increase in fission barrier height. It is seen in
Fig.~\ref{fig2} that the multi-qp states have wider and higher
fission barriers than the g.s., implying enhanced stability
against fission due to unpaired nucleons. This is consistent with
the observed very small spontaneous fission branch of
$\approx10^{-4}$ for the two isomers in
$^{254}$No~\cite{HessEPJA10}.

\begin{figure}
\includegraphics[scale=0.48]{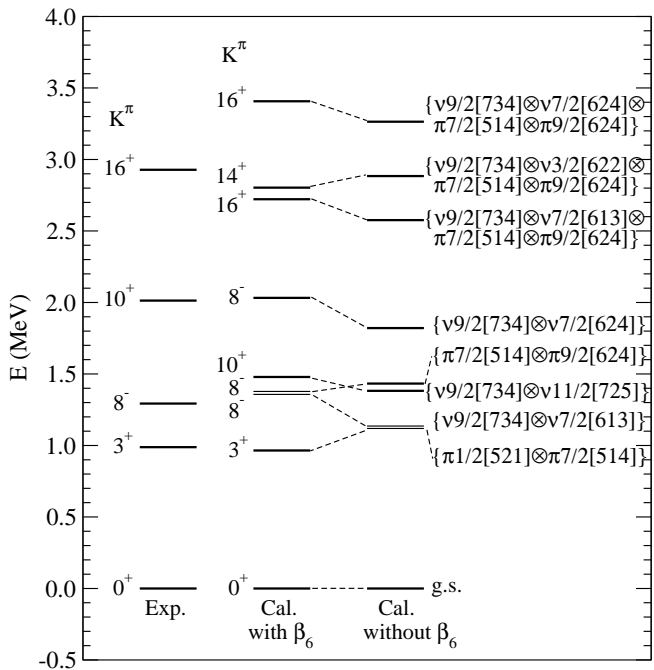}
\caption{\label{fig3}Calculations of $^{254}$No multi-qp states
with and without $\beta_6$ deformation, compared with experimental
data~\cite{HerzNature06,ClarkPLB10}.}
\end{figure}

The multi-qp states calculated with and without $\beta_6$
deformation are compared with experimental data in
Fig.~\ref{fig3}. (Note: since the excitation energy data for the
$K^\pi=3^+, 8^-$ states from different
experiments~\cite{HerzNature06,TanPRL06,KonND07,HessEPJA10,ClarkPLB10} are
similar, we adopt the earliest accurate data~\cite{HerzNature06};
the detailed data from the most recent
experiment~\cite{ClarkPLB10} are used for the other states.) The
$K^\pi=3^+$ state is firmly assigned the proton two-qp
configuration $\pi1/2^-[521]\otimes\pi7/2^-[514]$ through $g$
factor measurement~\cite{HerzNature06,TanPRL06,ClarkPLB10}. The
$K=3$ coupling is energetically favored over the $K=4$ coupling
due to the residual spin-spin interaction between the
quasiparticles~\cite{GalPR58,GalPR62}. According to the
Gallagher-Moszkowski (GM) rule~\cite{GalPR58,GalPR62}, the
spin-antiparallel coupling is energetically favored for two
quasineutrons or two quasiprotons, while the spin-parallel
coupling is lower in energy for the combination of a quasineutron
and a quasiproton. The splitting energies for the $A\approx180$
nuclei are found to be in the range of $\approx100-400$
keV~\cite{JainNPA95}. The energy is too small to substantially
change the calculation of a multi-qp state. Our model in its
present version does not include the residual spin-spin
interaction. The calculations usually well reproduce the
energetically favored coupling (see e.g.
Refs.~\cite{XuPRL04,XuPLB98}).

\begin{figure}
\includegraphics[scale=0.45]{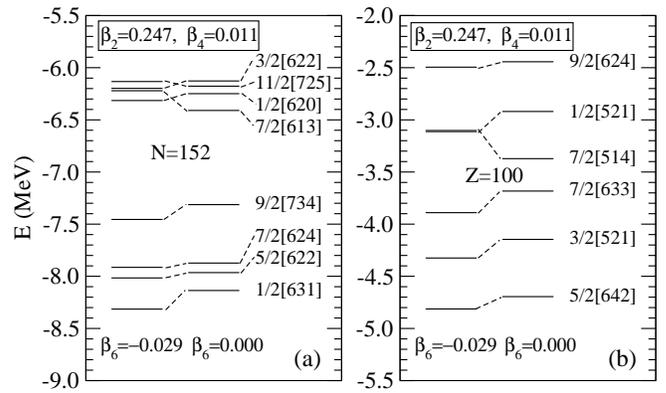}
\caption{\label{fig4}$^{254}$No neutron (a) and proton (b)
single-particle levels calculated using the Woods-Saxon potential
with the universal parameter set.}
\end{figure}

Our calculation of the $\pi1/2^-[521]\otimes\pi7/2^-[514]$
configuration with $\beta_6$ deformation gives an excitation
energy of 0.965 MeV, in very good agreement with the experimental
data 0.988 MeV~\cite{HerzNature06}. The low excitation energy
implies that the $\pi1/2^-[521]$ and $\pi7/2^-[514]$ orbits must
be close in energy. In Fig.~\ref{fig4}, we present the
single-particle levels calculated with and without $\beta_6$
deformation. One can see in Fig.~\ref{fig4} that the two orbits
become nearly degenerate due to $\beta_6$ deformation so that we
obtain an improved reproduction of the state with the inclusion of
the high order deformation. It is worth noting that $\beta_6$
deformation leads to an enlarged $Z=100$ deformed shell gap,
consistent with that predicted in Ref.~\cite{PatNPA91}.
Experiment~\cite{GrePRC08} has confirmed the existence of the gap
together with the stronger $N=152$ gap. The $K^\pi=3^+$ state is
of special interest because the $\pi1/2^-[521]$ orbit originates
from the spherical orbit $2f_{5/2}$ whose position relative to the
spin-orbit partner $2f_{7/2}$ determines whether $Z=114$ is a
magic number for the ``island of stability''. The good agreement
between experiments and our calculations with $\beta_6$
deformation demonstrates the importance of the high order
deformation in very heavy nuclei and the validity of the
Woods-Saxon potential in this mass region.

Unlike the $K^\pi=3^+$ state with its configuration unambiguously
assigned, the observed 266 ms $K^\pi=8^-$ isomer has its
configuration controversially assigned in the literature. The
proton two-qp configuration $\pi7/2^-[514]\otimes\pi9/2^+[624]$ is
suggested for the isomer in
Refs.~\cite{HerzNature06,TanPRL06,HessEPJA10}, while the most
recent experiment~\cite{ClarkPLB10} favors a neutron two-qp
configuration. There are two possible $K^\pi=8^-$ neutron two-qp
configurations, $\nu9/2^-[734]\otimes\nu7/2^+[613]$ and
$\nu9/2^-[734]\otimes\nu7/2^+[624]$. Our calculation of the latter
indicates that the state is too high in energy to be the isomer.
The high energy is because both orbits lie below the large $N=152$
shell gap. Therefore, it requires two neutrons to cross the gap to
form the state. The configuration favors the formation of an
isomer in $N=150$ nuclei where the Fermi surface is between the
two orbits. Indeed, low energy isomers with this configuration
were systematically observed in $N=150$
isotones~\cite{HerzPPNP08}. For the other $K^\pi=8^-$ neutron
two-qp configuration, $\nu9/2^-[734]\otimes\nu7/2^+[613]$, the
energy calculated with $\beta_6$ deformation is very similar to
that of the proton two-qp configuration
$\pi7/2^-[514]\otimes\pi9/2^+[624]$ (see Fig.~\ref{fig3}). Both
the calculated $K^\pi=8^-$ states are in better agreement with
experiments than those calculated without $\beta_6$ deformation.
This is attributed to the $\beta_6$ deformation that enhances the
$N=152$ and $Z=100$ deformed shell gaps, leading to increased
separation of the $\nu9/2^-[734]$ and $\nu7/2^+[613]$ orbits and
decreased separation of the $\pi7/2^-[514]$ and $\pi9/2^+[624]$
orbits. It should be noted that the $K=8$ coupling for the neutron
two-qp configuration is not the energetically favored one of the
GM doublet. When considering the residual spin-spin interaction,
the proton two-qp state, instead of the neutron two-qp state,
could be the lowest $K^\pi=8^-$ state. Nevertheless, they remain
close to each other because the GM splitting energy is small.
Experimental information such as the $g$ factor is needed to
distinguish between the two configurations for the $K^\pi=8^-$
isomer.

The two low-energy $K^\pi=8^-$ configurations can couple to form a
four-qp $K^\pi=16^+$ state, analogous to the well-known
$K^\pi=16^+$ isomer in $^{178}$Hf~\cite{WalNature99}. Indeed, a
four-qp 184 $\mu$s isomer has been observed. However, its
configuration is less clear than those of the two-qp states. Two
possible configurations,
$K^\pi=16^+\{\nu9/2^-[734]\otimes\nu7/2^+[624]\otimes\pi7/2^-
[514]\otimes\pi9/2^+[624]\}$ and
$K^\pi=14^+\{\nu9/2^-[734]\otimes\nu3/2^+[622]\otimes\pi7/2^-
[514]\otimes\pi9/2^+[624]\}$, were suggested in
Ref.~\cite{HerzNature06} and Refs.~\cite{TanPRL06,KonND07}, respectively.
The most recent experiment~\cite{ClarkPLB10} preferred a
spin-parity assignment of $K^\pi=16^+$. Our calculations shown in
Fig.~\ref{fig3} indicate that the configuration suggested in
Ref.~\cite{HerzNature06} is much higher than the four-qp
$K^\pi=16^+$ configuration involving the $\nu7/2^+[613]$ orbit.
This is due to the high energy of the
$\nu9/2^-[734]\otimes\nu7/2^+[624]$ coupling, as discussed above.
The $K^\pi=14^+$ configuration with a low-$\Omega$ orbit
$\nu3/2^+[622]$ involved is calculated to be also higher than the
$\nu9/2^-[734]\otimes\nu7/2^+[613]\otimes\pi7/2^-[514]\otimes\pi9/2^+[624]$
configuration. Consequently, the calculated lowest-lying
$K^\pi=16^+$ state is likely the 184 $\mu$s isomer due to its low
energy and high $K$ value, compatible with the experimental
evidence of a $K^\pi=16^+$ spin-parity
assignment~\cite{ClarkPLB10}. The excitation energy calculated
with $\beta_6$ deformation is 2.722 MeV, which is close to the
measured value of 2.928 MeV~\cite{ClarkPLB10}. In Fig.~\ref{fig3},
it can be seen that the inclusion of $\beta_6$ deformation
increases the calculated energy, making it closer to the
experimental value. Furthermore, the neutron component of
unfavored residual interaction is expected to further increase the
energy.

\begin{table}
\begin{threeparttable}
\caption{\label{tab1}Theoretical deformations and excitation
energies of multi-qp states in $^{254}$No.}
\begin{ruledtabular}
\begin{tabular}{cccccc}
 $K^\pi$ & Configuration\tnote{\dag} & $\beta_2$ & $\beta_4$ &
 $\beta_6$ & $E_x$(keV) \\
\hline
 $0^+$ & g.s. & 0.247 & 0.011 & -0.029 & 0  \\
 $3^+$ & ab & 0.247 & 0.011 & -0.030 & 965  \\
 $8^-$ & AB & 0.241 & 0.012 & -0.024 & 1357  \\
 $8^-$ & bc & 0.245 & 0.009 & -0.028 & 1378 \\
 $6^-$ & AE & 0.247 & 0.010 & -0.029 & 1427  \\
 $10^+$ & AD & 0.244 & 0.010 & -0.027 & 1479 \\
 $7^-$ & bd & 0.246 & 0.010 & -0.028 & 1481  \\
 $8^+$ & cd & 0.244 & 0.009 & -0.027 & 1658  \\
 $7^+$ & BC & 0.242 & 0.014 & -0.026 & 1774  \\
 $9^-$ & CD & 0.246 & 0.012 & -0.028 & 1881  \\
 $8^-$ & AC & 0.243 & 0.014 & -0.025 & 2032  \\
 $9^-$ & BD & 0.238 & 0.010 & -0.022 & 2237  \\
 $16^+$ & ABbc & 0.240 & 0.010 & -0.024 & 2722  \\
 $14^+$ & AEbc & 0.245 & 0.008 & -0.028 & 2803  \\
 $18^-$ & ADbc & 0.242 & 0.008 & -0.026 & 2845  \\
 $17^+$ & ABCD & 0.239 & 0.013 & -0.023 & 3158  \\
 $16^+$ & ACbc & 0.241 & 0.012 & -0.025 & 3407  \\
 $25^-$ & ABCDbc & 0.238 & 0.011 & -0.023 & 4522  \\
 $24^-$ & ABCDbd & 0.239 & 0.013 & -0.023 & 4631  \\
 $25^+$ & ABCDcd & 0.236 & 0.011 & -0.021 & 4774  \\
\end{tabular}
\begin{tablenotes}
\footnotesize \item[\dag] Neutron orbits $9/2^-$[734],
$7/2^+$[613], $7/2^+$[624], $11/2^-$[725], and $3/2^+$[622] are
represented by A, B, C, D, and E respectively. Proton orbits
$1/2^-$[521], $7/2^-$[514], $9/2^+$[624], and $7/2^+$[633] are
represented by a, b, c, and d respectively.
\end{tablenotes}
\end{ruledtabular}
\end{threeparttable}
\end{table}

In addition to all the multi-qp states observed before, a two-qp
$K^\pi=10^+$ state was observed in the most recent
experiment~\cite{ClarkPLB10}, with the configuration
$\nu9/2^-[734]\otimes\nu11/2^-[725]$ suggested. Fig.~\ref{fig3}
shows that the calculated excitation energy is 1.479 MeV, much
lower than the experimental data 2.013 MeV~\cite{ClarkPLB10}.
However, the $K^\pi=10^+$ state has unfavored spin-spin coupling
that would increase the excitation energy. The energy increment
could reach $\approx400$ keV as our calculated excitation energy
can be taken as the value for the favored coupling.

As shown in Fig.~\ref{fig4}, there exist several high-$\Omega$
orbits around the $^{254}$No Fermi surface that can couple to many
other high-$K$ states. Table~\ref{tab1} summarizes the
calculations with the inclusion of $\beta_6$ deformation. The
calculated excitation energy of the six-qp $K^\pi=25^-$ state is
4.522 MeV, comparable to 3.942 MeV, the excitation energy of the
observed $24^+$ g.s.~band member~\cite{HerzNature06}. The
$K^\pi=25^-$ state could be close to the yrast line (where the
state has the lowest energy among the states with the same angular
momentum), possibly forming an isomeric state.

In summary, the effects of the high order deformation, $\beta_6$,
on the high-$K$ isomers in $^{254}$No are investigated by applying
configuration-constrained PES calculations in $(\beta_2, \beta_4,
\beta_6)$ deformation space. The isomers gain extra binding energy
due to the $\beta_6$ deformation, implying enhanced stability
against fission. The high order deformation rearranges the
single-particle levels, leading to strengthened deformed shell
gaps at $N=152$ and $Z=100$, which influences the properties of
the multi-qp states. These effects are found to be significant.
All the observed multi-qp states in $^{254}$No are better
reproduced by the calculations with $\beta_6$ deformation. This
indicates the importance of the high order deformation in
calculating multi-qp states in very heavy nuclei.

We are grateful to T.L. Khoo and F.G. Kondev for suggesting the
present work. This work was supported in part by the US DOE under
Grants DE-FG02-08ER41533 and DE-FC02-07ER41457 (UNEDF, SciDAC-2),
and the Research Corporation; the Chinese Major State Basic
Research Development Program under Grant 2007CB815000; the
National Natural Science Foundation of China under Grants 10735010
and 10975006; and STFC and AWE plc (UK).

\end{document}